\documentclass[prd,preprint,superscriptaddress,amsmath,amssymb,nofootinbib]{revtex4}
\usepackage{graphicx}
\usepackage{dcolumn}
\usepackage{bm}
\usepackage{amssymb}
\usepackage{amsmath}
\usepackage{epsfig}    
\usepackage{color}
\usepackage{slashed}
\usepackage{hhline}
\usepackage{bbm}
\usepackage{float}
\usepackage{tikz-feynman}
\tikzfeynmanset{compat=1.1.0, warn luatex=false}

\def\be{\begin{equation}}
\def\ee{\end{equation}}
\newcommand{\bea}{\begin{eqnarray}}
\newcommand{\eea}{\end{eqnarray}}

\begin{document}

\preprint{
\begin{minipage}{5cm}
\small
\flushright
KYUSHU-HET-353
\end{minipage}} 

\title{Dynamical CP Violation from Non-Invertible Selection Rules}

\author{ Hiroshi Okada}
\email{hiroshi3okada@htu.edu.cn}
\affiliation{Department of Physics, Henan Normal University, Xinxiang 453007, China}

\author{Hajime Otsuka}
\email{otsuka.hajime@phys.kyushu-u.ac.jp}
\affiliation{Department of Physics, Kyushu University, 744 Motooka, Nishi-ku, Fukuoka 819-0395, Japan}
\affiliation{Quantum and Spacetime Research Institute (QuaSR), Kyushu University, 744 Motooka, Nishi-ku, Fukuoka, 819-0395, Japan}

\begin{abstract}
We propose a novel mechanism in which leptonic CP-violating phases are generated dynamically through the radiative breaking of non-invertible selection rules. 
In this framework, tree-level mass matrices, initially constrained by a CP-like symmetry within a non-invertible structure, acquire flavor-dependent phases once loop corrections are incorporated. 
Furthermore, these corrections can also generate mass terms, thereby addressing the mass hierarchy problem. 
As an illustrative example, we employ the Inverse Seesaw (ISS) model to demonstrate how the Majorana mass of the light sterile neutrino $N_L$ arises via this mechanism while simultaneously realizing CP violation. Although our analysis is carried out within the ISS framework, the mechanism has broader implications, potentially offering new perspectives on CP-related problems such as the strong CP problem, leptogenesis, and baryogenesis. This work thus establishes a foundation for exploring the dynamical breaking of non-invertible selection rules as a novel origin of CP violation in particle physics. 
 \end{abstract}

\date{\today}

\maketitle

\section{Introduction}
The Standard Model (SM) provides a remarkably successful framework for describing fundamental particles and their interactions. However, it fails to account for the experimentally established fact that neutrinos possess nonzero masses. This necessitates an extension of the SM. In constructing such an extension, we adopt the guiding principles that the relevant new physics should appear at the TeV scale, where it remains accessible to collider experiments and that large mass hierarchies should be avoided to prevent naturalness and minimize fine-tuning. 

Within this context, the Inverse Seesaw mechanism (ISS)~\cite{Mohapatra:1986bd, Wyler:1982dd} emerges as one of the most compelling candidates.
The ISS can be understood in terms of 't Hooft’s naturalness criterion: a parameter is naturally small if setting it to zero restores a symmetry of the theory. In this case, the smallness of neutrino masses arises from a tiny violation of lepton number symmetry. 
In the exact symmetry limit, neutrinos remain massless, while a small breaking of lepton number induces correspondingly small neutrino masses. 
This provides a natural explanation for the observed scale of neutrino masses without invoking unnaturally large hierarchies. 
Moreover, the mechanism is realizable at the TeV scale, placing it within the reach of current collider experiments and neutrino observatories, thereby offering a framework that is experimentally testable in the near future. 
In the minimal construction of the model, 
the relevant Lagrangian can be formulated as
\begin{align}
y_D \overline{L_L} (i\sigma_2) H^* N_R + M_D \overline{N_R} N_L +\delta\mu_L  \overline{N_L^C} N_L +{\rm h.c.},  \label{eq:iss_orig}
\end{align}
where $\sigma_2$ is the second Pauli matrix, $N_R(N_L)$ denotes right(left)-handed sterile neutrinos.
After spontaneous electroweak symmetry breaking with $\langle H\rangle\equiv[0,v_H/\sqrt2]^T$, where $v_H\equiv 246$ GeV, the Dirac mass term is defined as $m_D\equiv y_D v_H/\sqrt2$. 
Then, in the basis $[\nu_L,N^C_R,N_L]^T$, the neutral fermion mass matrix takes the  block form:
\begin{align}
\begin{pmatrix}
 0 & m_{D}^* &  0  \\
 m^\dag_{D} & 0 & M_D \\
0& M^T_D & \delta\mu_L 
  \end{pmatrix}. \label{eq:iss_mat}
\end{align}
Imposing the following mass hierarchy:
\begin{align}
\delta\mu_L \ll m_{D} \lesssim  M_D, \label{eq:mass-order}
\end{align}
the effective active neutrino mass matrix is obtained as
\begin{align}
m_\nu\simeq m^*_{D} (M_D^T)^{-1} \delta\mu_L M_D^{-1} m^\dag_{D}. \label{eq:iss}
\end{align}

The minimal Inverse Seesaw construction, while elegant, naturally raises several theoretical questions that must be addressed to provide a fully consistent framework:
\begin{itemize}
\item
{\it Realization of the Lagrangian}\\
The minimal model cannot be realized without invoking additional symmetries as explicitly found in Eq.~(\ref{eq:iss_orig}). In most scenarios, discrete or continuous symmetries, or through the introduction of new fields, are required to control the structure of the Lagrangian. Without such organizing principles, the model remains phenomenologically ad hoc.

\item {\it Explanation of the hierarchy ($\delta\mu_L\ll m_D\lesssim M_D$)}\\
The smallness of $\delta\mu_L$, particularly relative to $m_D$, demands a theoretical justification beyond ’t Hooft naturalness, which serves only as an interpretive guide rather than a constructive mechanism. Several approaches have been proposed, such as the radiative generation of $\delta\mu_L$, where loop corrections naturally suppress its magnitude~\cite{Nomura:2018ktz, Nomura:2024jxc, Nomura:2021adf, Okada:2012np, Jangid:2025krp}, or the use of higher-dimensional operators (e.g., dimension-five terms suppressed by a cutoff scale) which can yield a small effective $\delta\mu_L$~\cite{Okada:2012np, Abdallah:2011ew}. 
Both approaches typically require the imposition of additional symmetries to ensure consistency and stability of the hierarchy.

\item {\it Reduction of free parameters}\\
Like many seesaw-type models, the ISS introduces a large number of free parameters. Although flavor symmetries or other organizing principles can reduce this freedom, it is desirable to develop mechanisms that achieve parameter reduction without relying solely on conventional symmetry assumptions. 

\item {\it Origin of CP violation}\\
Understanding the origin of CP violation remains one of the central challenges in particle physics, with profound implications for phenomena such as the strong CP problem, leptogenesis, and baryogenesis, and its origin is not confined to the ISS model.
\end{itemize}

{\it Non-invertible Fusion Rules and Generalized CP}\\
Unlike conventional group-based symmetries, non-invertible selection rules do not originate from a group structure; instead, they are formally well-defined within the framework of hypergroups. These symmetries have been shown to possess unique utility in phenomenology. 
Their applications span a wide range of contexts, including quark–lepton texture analyses~\cite{Qu:2026omn, Kobayashi:2025ldi, Kobayashi:2024cvp, Kobayashi:2025znw,
Nomura:2025yoa}, dynamical breaking corrections~\cite{Suzuki:2025oov, Kobayashi:2025cwx, Nomura:2025sod,  Okada:2025kfm, Jangid:2025krp,
Jangid:2025thp, Nomura:2025tvz, Okada:2025adm, Okada:2026gxl, Chen:2025awz, Okada:2026iob}, strong CP physics~\cite{Liang:2025dkm,Kobayashi:2025thd, Kobayashi:2025rpx}, dark matter models~\cite{Suzuki:2025oov}, suppression of flavor changing neutral currents~\cite{Nakai:2025thw}, realization of family-independent matter symmetries~\cite{Kobayashi:2025lar}, and generalized CP constructions~\cite{Kobayashi:2025wty}.
The generalized CP framework enables a theory to be formulated with real parameters. If generalized CP holds exactly in the quark and lepton sectors, the Yukawa couplings can be treated as real at the tree level, and CP violation emerges through spontaneous CP breaking. 
This approach already offers partial progress toward reducing the number of free parameters in the theory. 
However, it was known that non-invertible fusion rules are, in general, violated by radiative corrections~\cite{Heckman:2024obe,Kaidi:2024wio,Funakoshi:2024uvy}. 
The true strength of non-invertible fusion rules lies in their capacity for dynamical breaking, i.e., a feature unattainable with conventional group-derived symmetries. This dynamical aspect is essential for realizing the full potential of these symmetries in model building.

{\it Application to the Inverse Seesaw}\\
In this work, we specifically consider $Z_3$ Tambara–Yamagami (TY) fusion rule~\cite{Dong:2025jra} having a CP-like symmetry within the Type-II Two-Higgs-Doublet Model~\cite{Branco:2011iw} extension of the ISS.~\footnote{Any invertible symmetries with gauged outer automorphism, for example, the $S_3$ gauging of $Z_3\times Z'_3$~\cite{Dong:2025jra}, could also realize our scenario.} 
 In this scenario:
\begin{itemize}
\item
At tree level, CP-like symmetry holds exactly in the visible sector, ensuring real Yukawa couplings. 
\item CP violation arises dynamically through the breaking of the TY fusion rule, appearing as a one-loop level correction. 
\item Crucially, the same dynamical breaking simultaneously generates the small parameter $\delta\mu_L$ at the one-loop level.
\end{itemize}
This dual emergence of $\delta\mu_L$ and CP violation from a single dynamical mechanism represents a significant conceptual advance. It provides a natural explanation for the hierarchy $\delta\mu_L\ll m_D, M_D$ while also addressing the proliferation of free parameters, thereby strengthening the theoretical foundation of the ISS.
In addition, the dynamical breaking mechanism naturally gives rise to a viable dark matter candidate in our model.

This paper is organized as follows. 
In section \ref{sec:selection-rule}, we review $Z_3$ TY fusion rule and its relation to generalized CP.
In section \ref{sec:iss}, we show our concrete model in the ISS framework.
Finally, section \ref{sec:con} is devoted to the summary and discussion.

\section{$Z_3$ Tambara-Yamagami fusion rule and generalized CP}
\label{sec:selection-rule}

Here, we briefly review a commutative fusion algebra with non-invertible fusion rules, in which fields are labeled by its basis elements. In particular, we focus on the $Z_3$ TY fusion algebra constructed from the finite set of basis elements $\{\mathbbm{I}, \mathbbm{a},\mathbbm{b}, \mathbbm{n} \}$:
\begin{align}
  \mathbbm{a}\otimes \mathbbm{b} = \mathbb{I},\quad
     \mathbbm{a} \otimes  \mathbbm{n} =  \mathbbm{b} \otimes  \mathbbm{n} = \mathbbm{n},\quad
  \mathbbm{n}\otimes \mathbbm{n} = \mathbb{I} +  \mathbbm{a} +  \mathbbm{b},
\end{align}
where $\mathbb{I}$ denotes the identity element. 
In the following analysis, we suppose that fields $\phi$ are labeled by these basis elements, while their conjugate fields $\phi^\ast$ are labeled by the corresponding inverse classes. 
Specifically, the inverse classes of $\mathbbm{n}$ and $\mathbb{I}$ coincide with themselves, but those of $\mathbbm{a}$ and $\mathbbm{b}$ are respectively given by $\mathbbm{b}$ and $\mathbbm{a}$. 
Such a structure can be naturally realized in string compactifications on orbifolds such as heterotic string theory on toroidal orbifolds~\cite{Dijkgraaf:1987vp,Kobayashi:2004ya,Kobayashi:2006wq,Beye:2014nxa,Thorngren:2021yso,Heckman:2024obe,Kaidi:2024wio,Kobayashi:2025ocp} and Calabi-Yau threefolds~\cite{Dong:2025pah,Dong:2026iwa}, and magnetized D-brane models in type IIB string theory~\cite{Kobayashi:2024yqq,Funakoshi:2024uvy}. In these settings, matter fields are labeled by conjugacy classes of a finite discrete group rather than by its representations. 
Since the multiplication rules of conjugacy classes differ
from those of group elements, they lead to non-invertible fusion rules, including the $Z_3$ TY rule, e.g., from the $S_3$ gauged conjugacy classes of the $\Delta(54)$ group (see Ref.~\cite{Dong:2025jra} for details). 

In addition, CP-like transformation can be consistently defined in field theories with non-invertible fusion rules. 
When a fusion algebra contains a $Z_2$ group-based symmetry relating $\phi$ and $\phi^\ast$, it corresponds to the charge conjugation. 
By combining the charge conjugation with the parity transformation, one can realize the CP-like transformation~\cite{Kobayashi:2025wty}. 
Such a $Z_2$ symmetry exists, e.g., in the $Z_3$ TY fusion algebra, corresponding to $\mathbbm{a}\leftrightarrow \mathbbm{b}$. 

Let us consider an interaction term constructed from fields labeled by $\mathbbm{a}$ or $\mathbbm{b}$\footnote{In this section, we disregard the Lorentz properties of scalar and spinor fields for the sake of simplicity.}:
\begin{align}
\label{eq:int}
    {\cal L} \supset g_n \phi_1\cdots \phi_n + g_n^\ast (\phi_1\cdots \phi_n)^\ast,
\end{align}
where $g_n$ denotes a coupling constant. 
Under CP-like transformation, an interaction term transforms as $g_n \phi_1^\ast \cdots \phi_n^\ast=g_n (\phi_1\cdots \phi_n)^\ast$, which is regarded as the Hermitian conjugate of $g_n^\ast \phi_1\cdots \phi_n$. 
Hence, the interaction term after the CP-like transformation is described by
\begin{align}
    g_n (\phi_1\cdots \phi_n)^\ast + g_n^\ast  \phi_1\cdots \phi_n.
\end{align}
When we impose CP invariance on the theory, the coupling constant $g_n$ is constrained to be real. 
This argument can be extended to a generic interaction term, regardless of whether fields are self-conjugate, as shown below.

Note that the CP-like transformation do not necessarily act on all basis elements. 
For instance, $\mathbbm{n}$ and $\mathbb{I}$ do not transform under the $Z_2$ transformation $\mathbbm{a}\leftrightarrow \mathbbm{b}$. 
Let us consider the following interaction term:
\begin{align}
\label{eq:int2}
    {\cal L} \supset f_n \phi_1\cdots \phi_n\Phi_1\cdots \Phi_m + f_n^\ast (\phi_1\cdots \phi_n)^\ast(\Phi_1\cdots \Phi_m)^\ast,
\end{align}
where $\phi_i$ is labeled by the basis element $\mathbbm{a}$ or $\mathbbm{b}$, whereas $\Phi_\ell$ is labeled by the other elements $\mathbbm{n}$ or $\mathbb{I}$. 
In this case, under the CP, the interaction term transforms as
\begin{align}
    f_n (\phi_1\cdots \phi_n)^\ast \Phi_1\cdots \Phi_m + f_n^\ast  \phi_1\cdots \phi_n (\Phi_1\cdots \Phi_m)^\ast.
\end{align}
Hence, a generic interaction between $\phi$ and $\Phi$ is not invariant under the CP-like transformation. 
However, when the field $\Phi$ belongs to a real representation under the CP, i.e., $\Phi^\ast=\Phi$, such as a real scalar and Majorana fermion, the coupling \eqref{eq:int2} is allowed when the coupling $f_n$ takes a real value. 
Furthermore, the interaction of $\Phi$ itself
\begin{align}
\label{eq:int3}
    {\cal L} \supset h_n \Phi_1\cdots \Phi_m + h_n^\ast (\Phi_1\cdots \Phi_m)^\ast,
\end{align}
is also invariant under the CP. The coupling $h_n$ is in general complex, but is restricted to be real for the case of a real scalar and single Majorana fermion.   
As a result, one can consider a theory invariant under the CP-like symmetry, even when the Lagrangian includes a complex coupling $h_n$. 
This is an essential point to realize a physical leptonic CP-violating phase through the dynamical violation of CP-like symmetry.

Here, we focus on the residual $Z_2$ symmetry of the fusion algebra, identified with charge conjugation. 
However, when the fusion algebra further contains group-based symmetries, including flavor symmetries, CP and flavor transformations do not commute in general, and one must consider a generalized CP transformation~\cite{Holthausen:2012dk,Chen:2014tpa,Kobayashi:2025wty}.

\section{Model setup}
\label{sec:iss}

We introduce a model with TY fusion rule. 
To construct the ISS, we introduce three families of neutral fermions $N_R$ and $N_L$ in addition to the SM framework as discussed in Introduction.

\begin{table}[H]
    \centering
\caption{Charge assignments of the fields under the $SU(2)_L \times U(1)_Y$ gauge symmetry and TY.}
\label{tab:1}
\begin{tabular}{|c||c|c|c|c|c|c||c|c|}\hline\hline  
& ~$\overline{L_L}$~& ~$\ell_R$~ & ~$N_L$~ & ~$N_R$~ & ~${H}$~ & ~$H'$~ & ~{$\chi_R$}~& ~{$S$}~    \\\hline\hline 
$SU(2)_L$   & $\bm{2}$  & $\bm{1}$  & $\bm{1}$ & $\bm{1}$ & $\bm{2}$   & $\bm{2}$    & $\bm{1}$ & $\bm{1}$  \\\hline 
$U(1)_Y$    & $\frac12$  & $-1$  & $0$  & $0$    & $\frac12$ & $-\frac12$    & $0$     & $0$    \\\hline
{TY}   & $ \mathbbm{a}$  & $  \mathbbm{a} $& $  \mathbbm{a}$ & $ \mathbbm{a}$ & $ \mathbbm{a}$  & $ \mathbbm{a}$ &  $\mathbbm{n}$& $\mathbbm{n}$ \\\hline 
 \end{tabular}
\end{table}

To realize the CP-invariant system in the visible sector, we need to assign $\mathbbm{a}$ under TY for all relevant particles; $\overline{L_L},\ \ell_R, N_L,\ N_R,\ H$, which are summarized in Table~\ref{tab:1}.\footnote{One can assign $\mathbbm{b}$ for all relevant particles, but the following discussion is the same with the case of $\mathbbm{a}$.} 
Then, allowed renormalizable terms are given by 
\begin{align}
y^\ell_{ij} \overline{L_{L_i}} H \ell_{R_j}
+
M_{D_{ab}}  \overline{N_{L_a}} N_{R_b} 
 +{\rm h.c.}, \label{eq:iss-1}
 \end{align}
which leads to $m_{\ell_{ij}}\equiv y^\ell_{ij} v_H/\sqrt2$ after the electroweak symmetry breaking. 
At this stage, however, only the $M_D$ term is allowed by this symmetry for the neutral fermion sector.
This is because neither $\overline{L_{L}} (i\sigma_2) H^* N_{R}$ and $\overline{L_{L}} (i\sigma_2) H^* N_{L}^C$ are forbidden by the TY fusion rule; $\mathbbm{I}\notin\mathbbm{a}\otimes\mathbbm{b}\otimes \mathbbm{b}$.
Hence, we introduce another isospin doublet Higgs $H'(\equiv [(v_{H'}+ h'+i z')/\sqrt{2},h'^-]^T)$ with $-1/2$ hypercharge and $\mathbbm{a}$ under TY. Then, we can write the following terms:
\begin{align}
y_{D_{ia}} \overline{L_{L_i}} H' N_{R_a} +{\rm h.c.}, \label{eq:iss-2}
 \end{align}
which provides us the Dirac term $m_D(\equiv v_{H'}/\sqrt2)$ after the electroweak symmetry breaking with
 $\sqrt{v_H^2+v_{H'}^2} \approx 246$ GeV.
{\it  Here, we would like to remind that nature of CP-like symmetry in the TY fusion algebra demands all the mass parameters to be real:
\begin{align}
\{ m_\ell,\ M_D,\ m_D \}\in{\rm Real}.\label{eq:real-a}
\end{align}}

It is worth noting that the vacuum expectation values $v_H$ and $v_{H'}$ are constrained to be real at the tree-level, protected by the TY fusion rule.~\footnote{Non-trivial terms such as $\delta\lambda(H'H)^2$ are generated at the one-loop level  through $(H'H)S^2$ where TY selection rule is broken. Since $\delta\lambda$ term is complex in general, $v_H$ or $v'_H$ can also be complex. However, we simply neglect this effect assuming it to be sufficiently small.}
This arises because the selection rule restricts all coefficients in the Higgs potential  to be real. 
Without loss of generality, one can rotate the particles such that two of $\{ m_\ell,\ M_D,\ m_D \}$ are diagonalized. For convenience in analyzing the lepton mass spectrum and mixing patterns, we adopt $m_\ell$ and $M_D$ are diagonal.

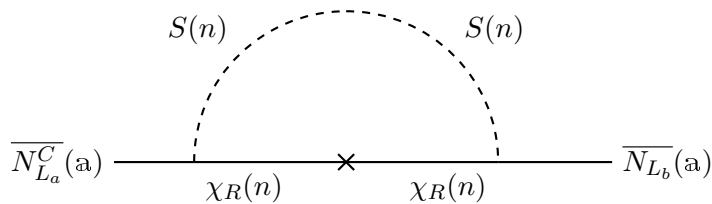
\begin{figure}[H]
\begin{center}
\begin{tikzpicture}
\begin{feynman}[large]
\vertex (a1) {\(\overline{ N_{L_a}^C} (\mathbbm{a})\)};
\vertex [right=1.8cm of a1] (b1);
\vertex [right=4.0cm of b1] (c1);
\vertex [right=1.5cm of c1] (d1) {\(\overline{N_{L_b}} (\mathbbm{a})\)};
\vertex [right=2.0cm of b1] (m1);
\vertex [above=2.0cm of m1] (e1);
\diagram [medium] {
(a1) -- (b1) -- [insertion={[size=3pt]0.5}, edge label'=\(\chi_R(n)~~~~~~~~~~~~~\chi_R(n)\)] (c1) -- (d1),
(b1) -- [scalar, quarter left, edge label=\(S(n)\)] (e1) -- [scalar, quarter left, edge label=\(S(n)\)] (c1),
};
\end{feynman}
\end{tikzpicture}
\end{center}
\vspace{-0.7cm}
\caption{Majorana mass matrices $m_L$ and $m_R$ at one-loop level. Here, {$Z_3$ TY} charges are given in parenthesis. One clearly finds that this one-loop dynamically violates {$Z_3$ TY} selection rule since $a\otimes a\neq {\mathbbm{1}}$ although each of vertex is invariant under {$Z_3$ TY}.}
\label{fig:Mass}
\end{figure}

The last task is to generate $\delta\mu_L$, which is prohibited at the tree-level. 
This term is arisen from help of three right-handed Majorana fermions $\chi_R$ and an inert singlet real scalar $S$. 
$\chi_R$ and $S$ are labeled by $\mathbbm{n}$ under TY which is self-conjugate algebra, and they correspond to $\Phi$ in the notation of Section~\ref{sec:selection-rule}. 
Introducing these particles leads us to induce the following terms:
\begin{align}
f_{{\alpha a}} \overline{\chi_{R_\alpha}} N_{L_a} S 
+
g_{{a\alpha}} \overline{N_{R_a}} \chi^C_{R_\alpha} S
+
M_{\chi_{\alpha}}  \overline{\chi_{R_\alpha}^C} \chi_{R_\alpha}
+
y_{S_{\alpha \beta}}  \overline{\chi_{R_\alpha}^C} \chi_{R_\alpha} S
+{\rm h.c.}, \label{eq:m_RL}
 \end{align}
 where $M_{X} $ is diagonal without loss of generality.
{\it Here, $f_{\alpha a}$ and $g_{\alpha a}$ are enforced to be real by analogy with \{$y^\ell,M_D,y_D\}$ in Eq.~(\ref{eq:real-a}), but $M_{\chi_\alpha}$ and $y_{S_{\alpha \beta}}$ can be treated as complex parameters.} 
The potential of $S$ is given by the standard form:
\begin{align}
    V= \mu_S^2 S^2 + \lambda S^4,
\end{align}
where a trilinear coupling is prohibited by the TY fusion rule. Throughout our analysis, we assume that $S$ does not develop a vacuum expectation value, maintaining $\langle S \rangle = 0$. 
Thus, the mass of $S$ is simply given by $m^2_S\equiv \mu^2_S+\frac{\lambda_{HS}}2 v_H^2$, where $\lambda_{HS}$ denotes the quartic coupling of $|H|^2|S|^2$.
Then, the Majorana mass term $\delta\mu_{L} \overline{N^C_L} N_L$ is generated at one-loop levels as shown in Fig.~\ref{fig:Mass}, and the resultant formula is found as~\footnote{Although $\delta\mu_{R} \overline{N_R} N^C_R$ is simultaneously generated at one-loop level, it does not contribute to the neutrino mass matrix. Therefore, we will not consider this term further in our discussion.}
 \begin{align}
\delta\mu_{L_{ab}}&=
\frac1{(4\pi)^2}
\sum_{\alpha=1}^3
f^T_{b\alpha} |M_{\chi_\alpha}| f_{\alpha a}
\frac{(1-r_S)r_\alpha\ln(r_\alpha)-(1-r_\alpha)r_S\ln(r_S)}{(r_S-r_\alpha)(1-r_\alpha)},  \label{eq:m_lR-form}
\end{align}
where $r_{S}\equiv m_S^2/\Lambda^2$ and $r_{\alpha}\equiv |M_{\chi_\alpha}|^2/\Lambda^2$ with $\Lambda$ being a cut-off scale.

It is noteworthy that computing Eq.~(\ref{eq:m_lR-form}) is more convenient in the real basis of the mass matrix $M_\chi$, even though
the three physical phases originally stem from $M_\chi$. 
First, we define the mass eigenvalues as $M_{\chi_\alpha}\equiv |M_{\chi_\alpha}| e^{2i\theta_\alpha}$ ($\alpha=1-3$).
The phases $\theta_\alpha$ can then be absorbed by the following field redefinition $\chi_{R_\alpha} \to  e^{-i\theta_\alpha} \chi_{R_\alpha}$.
While the coupling  $f$ is initially real, it acquires phases through the field redefinition of $\chi_R$.
This redefinition, in turn, shifts the coupling $f_{\alpha a}$ such that $f_{\alpha a}\to  e^{i\theta_\alpha} f_{\alpha a}$. 
Consequently, these three phases $\theta_{1,2,3}$ appear in $f$, and corresponds to the three physical phases: Dirac CP $\delta_{CP}$, two Majorana phases $\alpha_2$ and $\alpha_3$.
Here, we briefly discuss dependence on the scale of cut-off scale.
Once we fix $f=0.1$, $m_S=m_h/2$ GeV~\cite{Kanemura:2010sh} with $m_h\simeq 125.20$ GeV being the SM Higgs mass~\cite{ParticleDataGroup:2024cfk} 
and $|M_\chi|=1000$ GeV,
one finds the following values for different $\delta \mu_L$:
\begin{align}
&|\delta \mu_L|\simeq 0.58\,{\rm GeV} \qquad \mathrm{for}\,\, \Lambda=10^5\ {\rm GeV}, 
\label{eq:cut1}
\\ 
&|\delta \mu_L|\simeq 4.37\,{\rm GeV} \qquad \mathrm{for}\,\, \Lambda=10^{18}\ {\rm GeV}.
\label{eq:cut2}
\end{align}
It suggests that our theory works well up to the Planck scale,
and the cut-off scale dependence is not so large.

{\it Again, we would like to remind that $\delta\mu_L$ 
is not real anymore because of $M_{\chi_\alpha}$ which originates from terms with $\mathbbm{n}$.} Hence, the leptonic CP-violating phase can be dynamically generated through the radiative breaking of non-invertible selection rule.

We now turn to the discussion of the dark matter candidates, $\chi_R$ or $S$.
These particles are stabilized by the non-invertible element $\mathbbm{n}$ under the TY. 
Specifically, the lightest particle carrying the $\mathbbm{n}$ label serves as a viable dark matter candidate, %
as its stability is naturally ensured by the fusion rule $\mathbbm{n}\otimes \mathbbm{n}  = \mathbb{I}+  \mathbbm{a} +  \mathbbm{b}$, which contains the identity element.
However, $\chi_R$ does not have any interactions to explain the correct relic density. 
Thus, $S$ is the only channel to be the main source of the dark matter whose analysis in details has already been discussed in ref.~\cite{Kanemura:2010sh}.
It suggests that the allowed region to satisfy the observed relic density and bound on direct detection searches is at nearby half of the SM Higgs (or the second Higgs) mass GeV.

Now that we have successfully generated all the mass terms in Eq.~(\ref{eq:iss_mat}) with appropriate mass hierarchies in Eq.~(\ref{eq:mass-order}), the ISS formula for neutrino mass matrix is given as follows:
\begin{align}
m_\nu& \simeq m_{D} (M_D^T)^{-1} \delta\mu_L M_D^{-1} m^T_{D}.
 \label{eq:iss_ours}
\end{align}
Since we choose $m_D$ is diagonal, the mixing patterns arise from ${\cal W}\equiv (M_D^T)^{-1} \delta\mu_L M_D^{-1}$ with ${\cal W}$ being a complex symmetric $3\times3$ matrix.
Then, ${\cal W}$ can be diagonalized by a single unitary matrix $U_S$ as $D^{-1}_S\equiv U^T_S {\cal W} U_S$.
Therefore, $m_\nu$ can be rewritten by
\begin{align}
m_\nu = m_{D} U^*_S D_S^{-1} U_S^\dag m_D^T.
\end{align}
On the other hand, $m_{\nu}$ is also diagonalized by a single unitary matrix $U_{\nu}$ via $D_{\nu} \equiv U_{\nu}^{\dag} m_{\nu} U_{\nu}^*$.
Since we choose $m_\ell$ is diagonal, $U_\nu$ is identified with observed lepton mixing matrix~$U_{\rm PMNS}$~\cite{Maki:1962mu}.
Therefore, we can parametrize $m_D$ in terms of neutrino experimental results and some degrees of freedom in our model as
\begin{align}
m_D = U_{\rm PMNS} D^{1/2}_{\nu} {\cal O}_f D_S^{1/2} U_S^T \, ,
\label{eq:fExp}
\end{align}
where ${\cal O}_f$ is a $3\times 3$ complex orthogonal matrix; ${\cal O}_f^T {\cal O}_f = {\cal O}_f {\cal O}_f^T = \mathbbm{1}$.

The neutrino mass eigenstates can be rewritten in terms of the lightest neutrino mass $D_{\nu_{1(3)}}$ for NH(IH) and two experimental results; $\Delta m^2_{\rm sol}=D^2_{\nu_2}-D^2_{\nu_1}$
and $\Delta m^2_{\rm atm}$:
\begin{align}
{\rm NH}:\quad \Delta m^2_{\rm atm}= D^2_{\nu_3}-D^2_{\nu_1},
\\
{\rm IH}:\quad \Delta m^2_{\rm atm}= D^2_{\nu_2}-D^2_{\nu_3}.
\end{align}
Therefore,
\begin{align}
&{\rm NH}: D^2_{\nu_3}=\Delta m^2_{\rm atm} + D^2_{\nu_1},
\quad 
 D^2_{\nu_2}=\Delta m^2_{\rm sol} + D^2_{\nu_1},\\
& {\rm IH}: D^2_{\nu_2}=\Delta m^2_{\rm atm} + D^2_{\nu_3},
\quad 
 D^2_{\nu_1}=\Delta m^2_{\rm atm} -\Delta m^2_{\rm sol}+ D^2_{\nu_3}.
\end{align}
There are several experimental constraints on the neutrino masses. 
First, the sum of neutrino masses, which is denoted by $\sum D_{\nu} = D_{\nu_1} + D_{\nu_2}+ D_{\nu_3}$, is estimated to be its upper bound by the minimal standard cosmological model with CMB data~\cite{Planck:2018vyg}:
\begin{align}
\sum D_{\nu} \le 120 \, {\rm meV} \, .
\label{eq:sum1}
\end{align}
Recently, more stringent constraint is reported by DESI Collaboration that is $\sum D_{\nu} \le 71 \, {\rm meV}$ at 95\% CL~\cite{DESI:2024hhd}.

Next, the effective mass for neutrinoless double beta decay $m_{ee} \equiv |\sum_{i} D_{\nu_i} ((U_{\rm PMNS})_{ei})^2|$ is given by
\begin{align}
m_{ee} = \left| D_{\nu_1} c^2_{12} c^2_{13} + D_{\nu_2} s^2_{12} c^2_{13} e^{i \alpha_{2}} + D_{\nu_3} s^2_{13}e^{i (\alpha_{3} - 2 \delta_{CP})} \right| \, ,
\label{eq:0nu2beta}
\end{align}
where $s_{12,23,13} \, (c_{12,23,13})$, which are short-hand notations $\sin\zeta_{12,23,13} \, (\cos\zeta_{12,23,13})$, are mixing angles. 
The KamLAND-Zen collaboration gives an upper bound on $m_{ee}$ at the 90\% confidence level (CL)~\cite{KamLAND-Zen:2024eml} as
\begin{align}
m_{ee} < (28-122) \, {\rm meV} \, .
\label{eq:mee_current}
\end{align}

The third one is on effective electron neutrino mass that is defined by $m_{\nu_e}\equiv \sum_{i}D_{\nu_i}^2 |U_{\rm PMNS_{ei}}|^2$.
It is model independent observable, and given by
\begin{align}
m_{\nu_e} = \sqrt{D_{\nu_1}^2 c^2_{13} c^2_{12} + D_{\nu_2}^2 c^2_{13} s^2_{12} + D_{\nu_3}^2 s^2_{13}} \, .
\label{eq:mnue-modelindep}
\end{align}
Its upper bound at 90\% CL from KATRIN~\cite{KATRIN:2024cdt} is given by
\begin{align}
m_{\nu_e} \le 450 \, {\rm meV} \, ,
\end{align}
which is weaker than the other constraints.

The last one comes from non-unitarity constraints and mass hierarchy between $M_D$ and $m_D$, which is determined by several experiments such as the effective Weinberg angle, SM W boson mass, several ratios of Z boson fermionic decays, invisible decay of Z, electroweak universality, measurements of Cabbibo-Kobayashi-Maskawa matrix, and lepton flavor violations~\cite{Fernandez-Martinez:2016lgt}.
These suggest that a stringent bound is given by~\cite{Agostinho:2017wfs, Das:2017ski}~\footnote{More precisely, the bound depends on the components of  $M_D^{-1} m^\dag_{D}$.}
\begin{align}
| M_D^{-1} m^T_{D}| \lesssim 4.90\times 10^{-3}. \label{eq:f} 
\end{align}

Instead of performing a full numerical analysis, we provide an order estimation to satisfy the experimental bounds.
Considering the constraints from non-unitarity, we adopt the scaling $|m_D/M_D|\sim 10^{-3}$.
It follows that $|\delta\mu_L|\sim 10^{-4}$ GeV is required to reproduce the neutrino mass scale of 0.1 eV.
This requirement suggests a coupling of $f\sim0.001$ for the mass scales $|M_\chi| \sim10^3$ GeV, therefore $|m_D|\sim 1$ GeV, discussed in Eqs.~(\ref{eq:cut1}) and ($\ref{eq:cut2}$).

\section{Summary and discussion}
\label{sec:con}
In this study, we have developed a novel concept of {\bf dynamical CP violation} by exploiting the distinctive properties of non-invertible selection rules, specifically the $Z_3$ Tambara-Yamagami fusion rule. 
This framework provides a new scenario in which CP violation can be realized without relying on spontaneous CP breaking.

The proposed method is both rather general and straightforward: to all fields in which CP invariance is required, i.e., those with interactions restricted to real parameters, we assign $\mathbbm{a}$ in the TY fusion rule. 
This ensures that all relevant interactions begin with real coefficients.
To induce CP violation, we introduce self-conjugate fields, specifically an inert singlet scalar $S$ and Majorana fermions $\chi_R$ in our model, to which we assign $\mathbbm{n}$ in the TY fusion rule. 
When CP-invariant fields interact with these self-conjugate fields, new effective interactions emerge as radiative corrections, thereby breaking CP.
Since these corrections originate from the dynamical breaking of non-invertible selection rule, we designate this mechanism as {\bf dynamical CP violation}. 
Moreover, when corrections generate mass terms, they can address or alleviate the mass hierarchy problem. 
Thus, our proposal simultaneously offers a novel mechanism for dynamical CP breaking and a potential solution to the hierarchy problem, corresponding to an unprecedented idea in the field. 

To illustrate this general framework, we have applied it to the inverse seesaw model.
In particular, we have demonstrated that the small Majorana mass of $N_L$ can be dynamically generated through the radiative breaking of non-invertible selection rule. 
Although our analysis focused on the inverse seesaw framework, the general mechanism presented here has broader applicability.
It may provide new perspectives on long-standing CP-related problems such as the strong CP problem, leptogenesis, and baryogenesis. 
These directions remain open for future investigation, and we hope that this work will encourage further exploration and collaboration within the research community.

\acknowledgments

This work was supported by Zhongyuan Talent (Talent Recruitment Series) Foreign Experts Project (H.Okada) and JSPS KAKENHI Grant Numbers JP25H01539 (H.Otsuka) and JP26K07087 (H.Otsuka).

\bibliography{references}

\end{document}